\documentclass[12pt]{article}
\usepackage{epsf}

\newcommand{\veck}{{\bf k}}

\newcommand{\vecx}{{\bf x}}       %(vette x)
\newcommand{\vecp}{{\bf p}}       %(impulsvector p)

\newcommand{\lag}{\langle}
\newcommand{\rag}{\rangle}
\newcommand{\al}{\alpha}
\newcommand{\bt}{\beta}
\newcommand{\gm}{\gamma}
\newcommand{\dl}{\delta}
\newcommand{\ep}{\epsilon}

\newcommand{\kp}{\kappa}
\newcommand{\lm}{\lambda}
\newcommand{\rh}{\rho}

\newcommand{\ph}{\phi}
\newcommand{\vr}{\varphi}

\newcommand{\om}{\omega}

\newcommand{\Gm}{\Gamma}

\newcommand{\Sg}{\Sigma}
\newcommand{\Dl}{\Delta}

\newcommand{\half}{\frac{1}{2}}

\newcommand{\Tr}{\mbox{Tr}\,}

\newcommand{\eela}[1]{\label{#1}\end{equation}}
\newcommand{\eeala}[1]{\label{#1}\end{eqnarray}}

\newcommand{\ageq}{\mbox{}^{\textstyle >}_{\textstyle\sim}}
\newcommand{\be}{\begin{eqnarray}}
\newcommand{\ee}{\end{eqnarray}}
\newcommand{\bea}{\begin{eqnarray}}
\newcommand{\eea}{\end{eqnarray}}

\begin{document}

\title{
\vskip -100pt
{\begin{normalsize}
\mbox{} \hfill ITFA-98-08\\
\mbox{} \hfill  September 1998\\
\vskip  70pt
\end{normalsize}}
High Temperature Behavior of the Chern-Simons Diffusion Rate in the 
1+1~D Abelian Higgs Model 
}
\author{
Wai Hung Tang\thanks{email: tang@phys.uva.nl}\mbox{ }
and
Jan Smit\thanks{email: jsmit@phys.uva.nl}\mbox{ }
\\
Institute of Theoretical Physics, University of Amsterdam\\
Valckenierstraat 65, 1018 XE Amsterdam, the Netherlands
}
%\date{}
\maketitle
\begin{abstract}
We give arguments that in the 1+1 dimensional abelian Higgs model
the classical approximation can be good for the leading high temperature 
behavior of real time processes.
The Chern-Simons diffusion rate (`sphaleron rate') is studied numerically
in this approximation.
New results at high temperature show a $T^{2/3}$ behavior of the rate
at sufficiently small lattice spacing.\\[0.5cm]


\end{abstract}

\clearpage

\section{Introduction}

In theories of baryogenesis the rate of sphaleron transitions plays an 
important role \cite{CoKaNe93,RuSha96}. 
This rate is hard to calculate analytically, especially at high
temperatures. Numerical simulations in the quantum theory
have to face the problem of a
complex `Boltzmann factor' associated with real time correlation
functions at finite temperature. For this reason a classical
approximation was proposed some time ago \cite{GriRu88}, which
has been tested in the abelian
Higgs model in 1+1 dimensions 
%\cite{GriRuSha89,BochFo91,SmTa94,KraPo94,FoKraPo94,SmTa95}. 
[4--9].
At low temperatures the rate was found 
\cite{GriRuSha89,KraPo94,SmTa94} to agree with a semiclassical analytical 
calculation \cite{BochTsi89}, while at high
temperatures the results have been somewhat confusing:
ref.\ \cite{FoKraPo94} argued for a $T^{2/3}$ behavior, while we 
argued \cite{SmTa95} for $T^2$ behavior. We found in fact $T^2$ behavior at 
nonzero lattice spacing which turned into $T$-independence in the limit of 
zero lattice spacing. 
In the mean time simulations of the physically relevant
SU(2) models in 3+1 dimensions have been improving
%\cite{Ambea,AmKra95,TaSm96,MoTu97,MoHuMu97,Mo98} 
[11--16]
and the understanding of the classical 
approximation has been steadily increasing
%\cite{BoMcLeSmil95,AaSm96,AaSm97,BuJa97,Ar97,TaSm97,BoLa97,ArYa97}.
[17--24].
It therefore appropriate to return to the abelian Higgs model and
try to assess the situation in 1+1 dimensions.
We give a more full account of our earlier work briefly reported in 
\cite{SmTa95}.

In sect.\ 2 we introduce the model and the rate of fermion
number violation and we discuss some of their
properties. The classical approximation is introduced in sect.\
3 and studied in perturbation theory. Sect.\ 4 and 5 give details on the 
numerical method, sects.\ 6 and 7 contain results on the rate
in the regimes of low and high temperatures and sect.\ 8 contains
our conclusions. 
Details on the simulation algorithm using
`Kramers equation' are in the appendix.

\section{Abelian-Higgs model}

The classical abelian-Higgs model is given by the action 
\be S=-\int {\rm d}^2 x\; \left[ \frac{1}{4g^2} F_{\mu\nu}F^{\mu\nu} +
(D_\mu \phi)^\ast D^\mu \phi + \mu^2 |\phi|^2 +\lambda |\phi|^4
\right]\,, \ee
 with $F_{\mu\nu}=\partial_\mu A_\nu-\partial_\nu A_\mu$ and $D_\mu
\phi= (\partial_\mu - i A_\mu) \phi $ (our metric is given by
$g_{00} = - g_{11} = -1$). We recall that $A_{\mu}$,
$\sqrt{\lambda}$ and $g$ have dimension of mass, 
while $\ph$ is dimensionless in 1+1 dimensions.
The role of a dimensionless
coupling is played by 
$\lm/|\mu^2|$,
at given $\xi = g^2/\lambda$.

The classical
model has a Higgs phase for $\mu^2 < 0$, where
$\phi$ gets a ground state expectation value $|\phi|=v/\sqrt{2}$,
$v^2= -\mu^2/\lambda$, with masses $m_\phi^2 = 2\lambda v^2$,
$m_A^2=g^2v^2$. In the euclidean version of
the quantum theory there is a Berezinsky-Kosterlitz-Thouless
phase transition in the scalar model at $g=0$ which 
is turned into a crossover at nonzero gauge coupling
\cite{JoKoSi79,IchMa94}. For a recent lattice study, see \cite{DiHe96}.
At nonzero temperature there is only a
crossover in any case because then the system is infinite 
in one dimension only.
%\cite{Do60}.

A toy model for the electroweak theory is obtained by coupling to
fermions,
\be
S_F=-\int {\rm d}^2 x\; \left[ \bar{\psi} \gamma^\mu (\partial_\mu+i
\half
A_\mu \gamma_5)\psi + y \bar{\psi} (\phi P_R + \phi^\ast P_L)\psi
\right] \;,
\ee
where $P_{L,R} =(1\mp \gamma_5)/2$ are the chiral projectors. In the
quantum theory the fermion current is anomalous,
\be
\partial_\mu \bar{\psi} i \gamma^\mu \psi=-q\;,
\ee
with $q$ the topological charge density $(\epsilon_{01}=+1)$
\be q=\frac{1}{4\pi}\epsilon^{\mu\nu} F_{\mu\nu}=\partial_\mu C^\mu \;,
\ee
where $C^\mu$ the Chern-Simons current
\be
C^\mu=\frac{1}{2\pi}\epsilon^{\mu\nu} A_\nu \;.
\ee
As a consequence a change in Chern-Simons number
\be
C=\int_0^L {\rm d}x\; C^0=-\frac{1}{2\pi}\int_0^L {\rm d}x\; A_1
\ee
is accompanied by a change in fermion number 
\be 
F=\int_0^L {\rm d}x \; \bar\psi i \gamma^0 \psi \;, 
\ee
such that
\be F(t)-F(0)=-[C(t)-C(0)]=-\int_0^t {\rm d}t'\;\int_0^L {\rm
d}x\;q(x,t')\;.  \ee
We take space to be a circle with circumference $L$.

The sphaleron rate $\Gamma$ can be identified from the diffusion of
Chern-Simons number \cite{KhlSha88}, 
\be 
\Gamma =\frac{1}{t} \left\langle \left[ C(t)-C(0) \right]^2
\right\rangle, \hspace{5mm} t\rightarrow \infty. 
\label{defdiff}
\ee
In the following we shall neglect the influence of the fermions on the
Bose fields. Our task is to evaluate the above real time correlation
function.

Transitions with $\Delta C$ of order $1$ have to cope with the
sphaleron barrier of energy $E_s=(2/3)v^2m_\phi$, for which $C=1/2$
\cite{BochShap87}. The sphaleron configuration can be written in the
form
$$
A_0 = 0,\;\;\; A_1 = \frac{\pi}{L},\;\;\;
\phi = \frac{v}{\sqrt{2}}\tanh(\frac{m_{\phi}}{2}\,y)
\exp(i\frac{\pi}{L}y-i\frac{\pi}{2}),
$$
where $y=x-L/2 \in (-L/2,L/2)$.
At relatively low temperatures $m_\phi^2 \ll T \ll E_s$ the rate
has been calculated semiclassically with the result \cite{BochTsi89}
\bea
F &\equiv& \frac{\Gamma}{m_\phi^2 L} 
= f(\xi)\sqrt{\frac{E_s}{T}} e^{-E_s/T}\;,\\
f(\xi) &=& \left[ \frac{3}{(2\pi)^3} (s+1) \frac{\Gamma(\alpha+s+1)
\Gamma(\alpha-s)}{\Gamma(\alpha+1)\Gamma(\alpha)} \right]^{1/2} \;,
\\
\alpha &=& \sqrt{2\xi}\;,\hspace{5mm} s=\frac{1}{2}
\left(-1+\sqrt{1+8\xi}\right) \;.
\label{eq:analy}
\eea

At high temperatures the rate is difficult to calculate
analytically. 
%For an attempt in four dimensions, see \cite{Phil}.
On dimensional grounds it seems reasonable to expect
that for temperatures $T$ larger than any mass scale, in particular $T
\gg E_s$,
\be
%\frac{\Gamma}{m_\phi^2 L}=\kappa(\xi)\frac{T^2}{v^4 m_\phi^2}\;, 
\frac{\Gamma}{L}=\kappa(\xi)(v^{-2} T)^2\;, 
\label{Tsq}
\ee
with dimensionless $\kappa$. This formula is analogous to 
dimensional $\Gamma/L^3 = \kappa (\alpha_W T)^4$ in 3+1 dimensions,
with $v^{-2}$ playing the role of the electroweak coupling
$\alpha_W$. However, the behavior (\ref{Tsq}) is not obvious, since
the original couplings $\lm$ and $g^2$ are dimensionful. At high
temperature the dimensional reduction approximation leads to
temperature appearing in the combinations $\lm T$ and $g^2 T$. Since these
have engineering dimension three in mass units, we may expect the
behavior
\be
\frac{\Gamma}{L}=\tilde\kappa(\xi) (\lm T)^{2/3},
\label{Ttwothird}
\ee
instead of (\ref{Tsq}).

\section{Classical approximation}
The rate $\Gm$ defined by (\ref{defdiff}) is a nonperturbative quantity
and one approach to its computation is by numerical
simulation. This is well-known to be difficult since the
effective Boltzmann factor in real time processes is complex:
%\be\hspace{-3mm}
\be
\langle [C(t)-C(0)]^2 \rangle = 
\frac{
\Tr e^{-H/T}\, \left[ e^{iHt}\, C(0)\, e^{-iHt} - C(0)\right]^2
}{
\Tr e^{-H/T}
}. \label{2}
\ee
To cope with the complex weights a classical approximation has
been introduced \cite{GriRu88,GriRuSha89} in which the quantum mechanical
expectation value (\ref{2}) is replaced by a classical
expression,
\be
\langle [C(t)-C(0)]^2 \rangle
=\frac{ \int D\vr D\pi\, e^{-H_{\rm eff}(\vr,\pi)/T}\,
\left[ C(\vr(t),\pi(t)) - C(\vr,\pi)\right]^2 }
{\int D\vr D\pi\, e^{-H_{\rm eff}(\vr,\pi)/T} }.
\label{defclas}
\ee
Here $\vr$ and $\pi$ denote generic canonical variables and
$\vr(t)$ and $\pi(t)$ are solutions of the classical Hamilton
equations with an effective hamiltonian $H_{\rm eff}$ and initial conditions
$\vr(0)=\vr$, $\pi(0)=\pi$. 
This approximation is used at high temperature, where 
the important low momentum modes have high occupation numbers.

The classical approximation has been studied in perturbation theory 
in 3+1 dimensions 
%\cite{BoMcLeSmil95,Ar97,AaSm96,AaSm97,BuJa97}.
[17--21].
For static quantities the situation is well understood, as this 
corresponds to dimensional reduction in the imaginary time formalism,
of a 4D theory to a 3D theory which can be renormalized in the usual way. 
For time
dependent quantities there are divergent effects in  
classical gauge theories (related to the physics of Landau damping) 
which are non-local and cannot be absorbed by
local counterterms. In the renormalizable quantum theory such effects 
correspond to loop momenta of order of the temperature -- the well 
known hard thermal loop effects.  
 
In 1+1 dimensions the UV-divergencies are less severe. 
We shall now study the classical approximation for the abelian Higgs model
in perturbation theory, using the imaginary time 
formalism and making an analytic continuation to real time 
at a suitable point. 
For simplicity we choose the classically symmetric phase $\mu^2 > 0$ for the
starting point of perturbation theory. As a consequence the bare 
$A$-mass is zero, which causes infrared divergences in the 
diagrams. Such infrared divergences are presumably an 
artefact of perturbation theory and we assume they are cured automatically
in a correct nonperturbative treatment.
Here we shall use brute force and introduce an infrared 
regulator mass $m_A$. This crude method breaks gauge invariance 
but it is adequate for our purpose of elucidating the classical approximation.

\begin{figure}[htb]
\epsfxsize 90mm
\centerline{\epsfbox{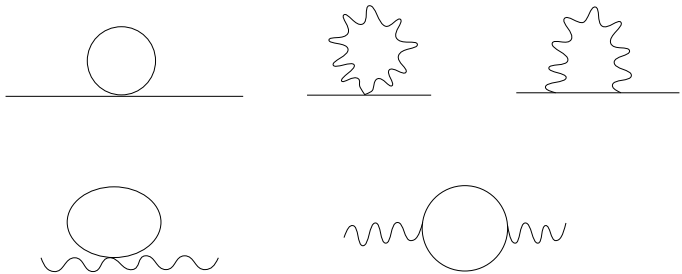}}
\caption[abh1]{Higgs and gauge field selfenergy diagrams.}
\label{abh:f1}
\end{figure}
Consider the 
gauge boson contribution to 
the $\ph$ selfenergy given by the appropriate diagrams in 
Fig.\ \ref{abh:f1}:
\bea
\Sg(p) &=& g^2 T\sum_n \int_{\veck}
\left[
\dl_{\mu\nu} 
- \frac{(k+2p)_{\mu} (k+2p)_{\nu}}{m_{\ph}^2 + (k+p)^2}
\right]
D_{\mu\nu}(k) ,
\label{Sg}\\
\int_{\veck}&\equiv &\int \frac{d^{D-1} k}{(2\pi)^{D-1}},
\eea
where $D$ ($\to 2$) is the number of spacetime dimensions, $k_D =  n 
2\pi T$ are the Matsubara frequencies
($n=0, \pm 1, \pm 2, \ldots$), and the vector propagator is given by
\be
D_{\mu\nu}(k) = \left[
\dl_{\mu\nu} + (\al-1) \frac{k_{\mu} k_{\nu}}{k^2}\right]
\frac{1}{m_A^2 + k^2}.
\ee
For illustration we first choose the Feynman gauge $\al = 1$ and
follow a standard method \cite{LeBe}, which consists of 
carrying out the sum over Matsubara modes in the loop diagrams 
and subsequently analytically continuing the time component $p_D$ of
the external momentum to real frequencies $p^0$.
Writing $(k +2p)^2 = 2[m_{\ph}^2 + (k+p)^2] - [m_A^2 + k^2] - 
(2m_{\ph}^2 - m_A^2 - 2p^2)$, to conveniently 
cancel the time component of the 
loop momentum in the numerator against the denominators, we get
\bea
\Sg &=& g^2 \int_{\veck}\left\{ 
(D-2)\frac{1 + 2n}{2\om} 
+ \frac{1 + 2n'}{2\om'} 
+ \frac{2m_{\ph}^2 - m_A^2 - 2\vecp^2 + 2p_0^2 }{2\om\om'}
\right.\nonumber\\
&&\left.
\times \left[
\frac{(1 + n + n')(\om + \om')}{(\om + \om')^2 - p^2_0}
- \frac{(n - n')(\om - \om')}{(\om - \om')^2 - p^2_0}
\right] \right\},
%\\
%\Sg_b &=& g^2 \int_{\veck} \frac{1 + 2n}{2\om},
\eea
where 
\be
n &=& \frac{1}{e^{\om/T} - 1},\;\;\; n' = \frac{1}{e^{\om'/T} - 1},
\;\;\; \nonumber\\
 \om &=& \sqrt{m_A^2 + \veck^2},
\;\;\; \om' = \sqrt{m_{\ph}^2 + (\veck + \vecp)^2}.
\label{nom}
\ee
%and $p^2 = \vecp^2 - p_0^2$.
%
The external real-time frequency $p^0 = -p_0$ is supposed to 
have a 
small positive imaginary part, $p^0 \to p^0 + i\ep$, which 
corresponds to retarded boundary conditions. 

The zero temperature 
contribution, obtained by letting $n,n' \to 0$, has a logarithmic 
UV-divergence which can be canceled by a mass counterterm. 
The temperature dependent contribution is finite as usual. 
We now add the mass counterterm to $\Sg$, 
assume dimensional renormalization in the MS-bar scheme, and denote
the resulting selfenergy by $\bar\Sg$. We are interested in its high
temperature behavior when $T \gg m_A$, $m_{\ph}$, $|\vecp|$, $p^0$. 

For high temperatures
%($T \gg m_A$, $m_{\ph}$, $|\vecp|$, $p^0$)
it turns out that  
the leading contribution in a high temperature expansion 
is obtained by substituting the leading behavior of
the Bose distribution functions $n$ and $n'$,
\be
\half + n \to \frac{T}{\om},
\;\;\; \half + n' \to \frac{T}{\om'}.
\label{nc}
\ee
The resulting expression can be written as
\bea
\Sg_c &=& g^2 T \int_{\veck} \left\{
\frac{D-2}{\om^2} + \frac{1}{\om'^2} + 
\frac{2m_{\ph}^2 - m_A^2 - 2\vecp^2 + 2p_0^2}{2\om^2 \om'^2}
\right.\nonumber\\
&&\left.\times \left[
\frac{(\om + \om')^2}{(\om + \om')^2 - p^2_0}
+ \frac{(\om - \om')^2}{(\om - \om')^2 - p^2_0}
\right] \right\}.
\label{Sgc}
\eea
This is finite for $D=2$, which justifies
the replacement (\ref{nc}) under the integral.
This contribution is actually identical to the classical 
selfenergy which follows from (\ref{defclas}), hence the subscript $c$.
Intuitively this is plausible, since 
large occupation numbers $n,n'$ imply classical behavior, although
it is of course not obvious that solving the classical equations of 
motion and averaging over initial conditions according to 
(\ref{defclas}) leads to the same correlation functions as
making the high temperature approximation (\ref{nc}). See
\cite{AaSm97} for more details in scalar field theory. 

{}For $p^0=0$ the classical selfenergy $\Sg_c$ goes over into
the selfenergy of the naively dimensionally reduced theory, the one dimensional
field theory obtained by keeping only the zero mode ($n=0$) in (\ref{Sg}).  
In this case $\Sg$ is the selfenergy of the 
zero mode $\int_0^{1/T} dx_D \ph(\vecx,x_D)$. 
%while the replacement (\ref{nc})
%is equivalent to keeping only the zero mode in (\ref{Sg}).

The difference between the classical and quantum selfenergy is 
subleading for large $T$ and it can be expressed as
\bea
\bar\Sg  - \Sg_c
&=& \frac{g^2}{2\pi}\,\left( c -\ln \frac{T}{\bar\nu} 
%- \ln\left(\frac{4\pi e^{-\gm_E} T}{\bar\nu}\right)^2
\right) 
+ O(T^{-1})
\label{deltaSg}
,\\
c&=& -\ln2 - 1 + \int_{0}^{\infty} dx\,
\left[
\left(1 + \frac{2}{e^x -1} -\frac{2}{x}\right) \frac{1}{x} 
-\frac{1}{(1+x)}\right],\nonumber
\eea
where $\bar\nu$ is the scale parameter in the MS-bar scheme.
We have subtracted and added the logarithmic UV-divergence in the form 
$\int_{\vecx} (1+|\vecx|)^{-1}$. The momentum dependence of 
$\bar\Sg  - \Sg_c$ is in the $O(T^{-1})$ terms. 
Although $\bar\Sg  - \Sg_c$ has been
calculated in the Feynman gauge, its leading $T$ contribution
given by the first term in (\ref{deltaSg}) is in fact independent of the
gauge parameter $\al$. This can be seen by going back to (\ref{Sg}),
setting $p=0$ and subtracting the $n=0$ mode. 
We can easily add the contribution of the $\lm(\ph^*\ph)^2$
interaction to the selfenergy. The complete one loop
$\bar\Sg  - \Sg_c$ is obtained by  $g^2 \to g^2 + 4\lm$ in (\ref{deltaSg}).

Consider next the gauge field self energy tensor $\Pi_{\mu\nu}$ corresponding
to the diagrams in Fig.\ 1,
\be
\Pi_{\mu\nu}(p) = g^2 T\sum_n\int_{\veck}
\left[
\frac{2\dl_{\mu\nu}}{m_{\ph}^2 + k^2} - 
\frac{(2k+p)_{\mu} (2k+p)_{\nu}}{m_{\ph}^2 + k^2)(m_{\ph}^2 + (k+p)^2)}
\right].
\ee
The logarithmic divergence of each of the two contributing diagrams cancels
in the sum, which can be shown by using the Ward-Takahashi identity
$p_{\mu} \Pi_{\mu\nu}(p) = 0$. In 1+1 dimensions this identity furthermore
tells us there is only one independent component in $\Pi_{\mu\nu}$,
which we choose to be $\Pi_{11}$. Summing over the Matsubara frequencies and
continuing $p_2$ analytically to $p^0$ we get
\bea
\Pi_{11} &=& g^2 \int_{\veck}
\left\{
\frac{2(1+2n)}{2\om} - \frac{(2k_1 + p_1)^2}{2\om\om'}
\right.\nonumber\\
&&\left.
\times \left[
\frac{(1 + n + n')(\om + \om')}{(\om + \om')^2 - p^2_0}
- \frac{(n - n')(\om - \om')}{(\om - \om')^2 - p^2_0}
\right] \right\},
\eea
where now $\om = \sqrt{m_{\ph}^2 + \veck^2}$ ($\om'$, $n$ and $n'$ are as
in (\ref{nom})).
The dominant large $T$ behavior of $\Pi_{11}$ is given by its classical
approximation obtained with the substitution (\ref{nc}),
\be
\Pi_{11}^c = g^2 T\int_{\veck} \left\{
\frac{2}{\om^2} - \frac{(2k_1 + p_1)^2}{2\om^2\om'^2}
%\right.\nonumber\\ &&\left.\times 
\left[
\frac{(\om + \om')^2}{(\om + \om')^2 - p^2_0}
+ \frac{(\om - \om')^2}{(\om - \om')^2 - p^2_0}
\right] \right\},
\label{Pic}
\ee
which is UV-finite and of order $T$. The next to leading behavior
turns out to be of order $T^0$ (not $\ln T$):
\be
\Pi_{11}- \Pi_{11}^c = \frac{g^2}{\pi}\, \frac{p_0^2}{p_1^2 - p_0^2}
+ O(T^{-1}).
\label{deltaPi}
\ee
This contribution corresponds to loop momenta $\veck$ of order $T$ and
it is therefore the analogue of the hard thermal loop expression in 3+1 
dimensions. Here in 1+1 dimensions it is subdominant to the classical
$\Pi_{11}^c$. 
All other diagrams are superficially UV-convergent in two dimensions
and for convergent diagrams
the corrections to the approximation (\ref{nc}) are down by
two powers $T$, as follows for example from the fact that they
correspond to the $n\neq 0$ terms in the summation over Matsubara
frequencies.

We conclude that in two dimensions the classical
theory is a good approximation to the quantum theory, for
weak coupling. The classical theory is UV-finite. To minimize
the difference between the classical and quantum theory we can
match the classical mass parameter $\mu_c^2$ of the Higgs field according
to
\be
\mu_c^2 = \bar\mu^2 + \bar\Sg  - \Sg_c = \bar\mu^2 +
\frac{g^2 + 4\lm}{2\pi}\,\left( c 
%- \ln\left(\frac{4\pi e^{-\gm_E} T}{\bar\nu}\right)^2
- \ln\frac{T}{\bar\nu}
\right) 
,
\label{mumatch}
\ee
where the bar on the right hand side indicates the MS-bar scheme.
%This matching depends on the strengths of the couplings and on the
%temperature. 
However, this matching is only of limited use since we cannot match
the corresponding non-analytic subdominant terms in $\Pi_{\mu\nu}$
with a local classical action. 

An important point is now that we can take a limit of weak coupling and
high temperature, such that the classical approximation becomes
exact: 
\be
\lm = v^{-2} |\bar\mu^2|,\;\;\; 
g^2 = \xi \lm,\;\;\;
T = v^2 |\bar\mu| T',\;\;\;
v^2 \to \infty.
\label{limit}
\ee 
The limit is such that $T'$ and $\xi$ are 
kept fixed. Note that $T' = \lm T/|\bar\mu|^3$, and $\lm T$ and $g^2 T$
are the combinations appearing in the dimensionally reduced theory.
In this limit $\mu_c^2/\bar\mu^2 \to 1$ and the quantum corrections
(\ref{deltaSg}) and (\ref{deltaPi}) go to zero.
In (\ref{limit}) we have written $|\mu|$ such that it also applies to 
the classical Higgs phase, for which $\mu^2 < 0$.
Instead of $\bar\mu$ we can of course use any
convenient mass scale, e.g.\ the zero temperature Higgs mass.

If we do not take the above limit, the question arises, how accurate is
the classical approximation. The above calculations are carried out
for external frequencies and momenta which are small compared to the 
temperature. 
On the other hand, in real time (as opposed to frequency space) we also
may expect a limited region of validity.
The correction (\ref{deltaPi}) leads to propagation
and situations might exist where it becomes substantial, depending on
the physical quantity under study. 
In the application to the computation of the sphaleron rate,
the Chern-Simons number is a zero momentum observable (explicitly so
in Coulomb gauge), and the rate is should be dominated by low frequencies
(cf.\ the cosine transform method in \cite{MoTu97}). 

The classical approximation is on a much better footing in 1+1 dimensions
than in 3+1 dimensions, 
%where the classical theory is linearly divergent. 
because it gives directly the dominant high temperature behavior of correlation 
functions, not only in the static case corresponding to
dimensional reduction, but also in case of time dependence.
(The classical partition function still
suffers from the Rayleigh-Einstein-Jeans divergence.)
%For static quantities the 3+1 situation is well understood, 
%as this corresponds to dimensional reduction in the imaginary time formalism
%of a 4D theory to a 3D theory, which can be renormalized.
In 4D, matching mass parameters between the original theory and the 
dimensionally reduced theory involves not only the classical combinations 
$g^2 T$ and $\lm T$, but also $g^2 T^2$ and 
$\lm T^2$. The latter combinations correspond to the 
well known hard thermal loops effects,
which are the result of loop momenta of order of the temperature, 
and which reflect the quadratic divergences in the bare theory.
Since $g^2 T^2$ dominates over $g^2 T|\bar\mu|$, the details of the
matching formulas are important in four dimensions even in the
weak coupling limit.
%
%However, for time dependent quantities the divergences of the
%classical theory lead to complications as they correspond to
%non-local behavior \cite{Ar97,AaSm97,ArYa97}.
In 2D, such matching can be ignored in the weak coupling limit
(\ref{limit}).

\section{Lattice regularization}

The classical partition function is
still a functional integral over all $\vr(\vecx)$ and $\pi(\vecx)$ and
needs regularization. In the numerical simulations this is
provided by a lattice, for which the action takes the form.
\bea 
S &=& \int dt\; a \sum_{n=0}^{N-1}
\left\{\frac{1}{2g^2} \left[({A_0}_{n+1} - {A_0}_n)/a -\partial_0
{A_1}_n \right]^2 \right.\nonumber\\ 
&& \mbox{} 
+ |(\partial_0-i {A_0}_n) \phi_n |^2 
-|\exp(-ia{A_1}_n) \phi_{n+1}-\phi_n|^2 /a^2 \nonumber\\ 
&& \left. \mbox{}
- \lambda \left[ |\phi_n|^2 - \frac{v^2}{2} \right]^2
\right\} \;, 
\eea
where $a$ is the lattice distance, $L=Na$ and  $v^2 = -\mu^2/\lm$. From
now on we only consider the case $\mu^2 < 0$.
%The parameters $g^2$,
%$\lambda$ and $v^2=-\mu^2/\lambda$ are now to be interpreted as {\em
%effective} parameters which depend in principle on the cutoff,
%temperature and coupling. The $\ldots$ denote corrections to the
%classical form shown. As discussed in the previous section, these
%corrections will be small for weak coupling and $aT/\pi$ not too large
%to small.
In the previous section we gave arguments that we can
use this action in the classical approximation, provided we replace
$\mu^2$ by an effective parameter $\mu_c^2$ which is related to
$\bar\mu^2$ in the MS-bar scheme by (\ref{mumatch}). In the weak
coupling-high temperature limit (\ref{limit}) the difference between
$\mu_c^2$ and $\bar\mu^2$ may be neglected, which is what we shall do
in the following. Accordingly, we drop the subscript $c$ on $\mu^2$
and $v^2$.

We use the Coulomb gauge to obtain a hamiltonian
description. The gauge condition ${A_1}_{n+1} - {A_1}_n=0$ states that
$A_1$ is independent of $n$ and related to the Chern-Simons number $C$
by,
\be
aA_1=2\pi C/N\;.
\ee
The Gauss constraint
\be 
\sum_n \Delta_{mn} {A^0}_n \equiv {A^0}_{m+1}+{A^0}_{m-1}-2{A^0}_m
= -a^2 g^2 {j^0}_m\;, 
\ee
where $j^0=-j_0$ is the charge density,
\be 
j_0=-i \phi^\ast D_0 \phi + i (D_0 \phi)^\ast \phi \;, 
\ee
is solved explicitly as 
\be
{A^0}_m=-a^2 g^2 \sum_n \Delta_{mn}^{-1} {j^0}_n,
\ee
under conditions with zero total charge $Q=a \sum_n
j_n^0$. The canonical momenta are defined by $\pi_n=\partial {\rm L}/
\partial \dot{\phi}_n$, etc. where $S=\int {\rm d}t\; {\rm L}$.

It will be convenient in the following to use $|\mu| = \sqrt{\lm} v$
as the basic unit of mass and to use scaled quantities,
indicated with a prime, 
\be 
a=\frac{a'}{v\sqrt{\lambda}}\;,\hspace{5mm}
t=\frac{t'}{v\sqrt{\lambda}}\;,\hspace{5mm} \phi=v
\phi'\;,\hspace{5mm} A_\mu=v\sqrt{\lambda} A_\mu'\;.
\label{prime}
\ee
Then
%
%\bea S_{\rm eff}(A,\phi) &=& v^2 S'(A',\phi')\;,\\
%
%S'(A',\phi') &=& S_{\rm eff}(A',\phi')_{\left|g^2\rightarrow \xi,\;
%\lambda\rightarrow 1,\; v^2\rightarrow 1,\; a\rightarrow a'\right.}
%\;.  \eea
\bea 
S &=& v^2 S',\\
S' &=& \int  dt'\; a' \sum_{n=0}^{N-1}
\left\{\frac{1}{2\xi} \left[({A'_0}_{n+1} - {A'_0}_n)/a' -\partial'_0
{A'_1}_n \right]^2 \right.\nonumber\\ 
&& \mbox{} 
+ |(\partial'_0-i {A'_0}_n) \phi'_n |^2 
-|\exp(-ia'{A'_1}_n) \phi'_{n+1}-\phi'_n|^2 /a'^2 \nonumber\\ 
&& \left. \mbox{} 
- \left[ |\phi'_n|^2 - \frac{1}{2} \right]^2
\right\} \;, 
\eea
and the hamiltonian can be written as
\bea
H &=& v^3\sqrt{\lambda} H'\;,\\
H' &=& \frac{1}{2} \frac{\xi L'}{(2\pi)^2} P_C^2 +\sum_{n=0}^{N-1} \left\{
\frac{|\pi_n'|^2}{a'} + \frac{|\exp(-i2\pi C/N)
\phi_{n+1}'-\phi_n'|^2}{a'} \right. \nonumber\\
&& \left. +a'\left(|\phi'_n|^2-\frac{1}{2}\right)^2 +\frac{a'^3 \xi}{2}
\sum_{m,n=0}^{N-1} j'^0_m \Delta_{mn}^{-1} j'^0_n \right\}\;,
\label{Hprime}\\
{j'}_n^0&=&(-i{\pi'}_n {\phi'}_n + i {\pi'}_n^\ast {\phi'}_n^\ast)/a'\;,
\eea
where $P_C$, ${\pi'}_n$ and ${\pi'}_n^\ast$ are the canonical momenta
conjugate to $C$, ${\phi'}_n$ and ${\phi'}_n^\ast$. 
The Poisson brackets are normalized to one, e.g.\
$\{\phi'_m,\pi'_n\}=\delta_{mn}$. 
The classical partition function is given by
\be 
Z=\int {\rm d}P_C\;{\rm d}C\; \prod_n ({\rm d}{\pi'}_n\;{\rm
d}{\pi'}_n^\ast\; {\rm d}{\phi'}_n\;{\rm d}{\phi'}_n^\ast)\;
\exp\left(-\frac{H'}{T'}\right) 
\ee
with $T'$ defined by
\be 
T=v^3 \sqrt{\lambda} T'\;, 
%\;\;\; T'=\lm T/|\mu|^3.
\ee
in accordance with (\ref{limit}).
The $C$ integration is over the compact domain 
$-N/2 \le C \le N/2$ (mod $N$).
We also define a scaled diffusion rate $\Gamma'$ by 
\be 
\Delta(t) \equiv \left\langle [C(t)-C(0)]^2 \right\rangle \rightarrow \Gamma
t=\Gamma' t'\;, 
\ee
for large times.

It is easy to see that there are $N$ classical ground states given by
$$
C = k = 0,1, \cdots, N-1,\;\;\;
\phi' = \frac{1}{\sqrt{2}}\, \exp(i 2\pi kn/N),
$$
and $P_C = \pi'_n = 0$. Writing
$\ph'_n = \rh_n \exp(i\theta_n)$ we can introduce a
winding number of the scalar field by
$$
w = \frac{1}{2\pi}\sum_{n=0}^{N-1} \partial \theta_n,\;\;\;
\partial \theta_n \equiv 
\theta_{n+1} -\theta_n\;\; \mbox{(mod $2\pi$)} \in (-\pi,\pi].
\label{winding}
$$
Then also $w=k$ in the ground states. Both $C$ and $w$ are defined modulo $N$.

For convenience we record various scaled quantities:
\bea 
E_s &=& v^3\sqrt{\lambda} E_s'\;,\hspace{3mm}
m_\phi=v\sqrt{\lambda} {m'}_\phi\;,\hspace{3mm}
m_A=v\sqrt{\lambda} {m'}_A\;, \\
E'_s &=& \frac{2\sqrt{2}}{3}\;,\hspace{3mm}
{m'}_\phi=\sqrt{2}\;,\hspace{3mm} {m'}_A=\sqrt{\xi}\;, 
\eea
which characterize the scaled theory given by the action $S'$ or
hamiltonian $H'$.
We also have
\bea
F &=& 
\frac{\Gamma}{m_{\phi}^2 L}
 =
\frac{\Gamma'}{m'^2_{\phi} L'}
\nonumber\\
&\rightarrow& \half\,\kappa T'^2
\;\;\mbox{or}\;\; \rightarrow \half\, \tilde \kappa T'^{2/3},
\label{alternatives}
\eea
for the large temperature behavior (\ref{Tsq}) or (\ref{Ttwothird}).

%In the following we shall understand all variables and parameters to be 
%`primed' and omit the primes unless the distinction is explicitly needed.

\section{Numerical method}

It is straightforward to derive Hamilton's equations from the Coulomb 
gauge hamiltonian (\ref{Hprime}). 
We use a second order Langevin procedure to generate initial conditions
according to the canonical ensemble. 
In this procedure, also known as the Kramers equation method,
Hamilton's equations are modified by adding noise and friction
terms to the equations for the canonical momenta. If this is done 
straightforwardly local `current conservation' 
$0=\partial_0 j^0_n + (j^1_{n+1} - j^1_n)/a \sim \partial_{\mu} j^{\mu}$
would be violated, since $j^0$ contains the momentum of the scalar field.
Consequently the total charge $Q$ would be nonzero, which is a problem for
the evaluation of the Coulomb energy.
To deal with these constraints (which are similar to the Gauss constraint
in the temporal gauge) we follow ref.~\cite{BochFo91} and use
polar coordinates $\ph'=\rh\exp(i\theta)$. The random Langevin
forces are then only applied
to the gauge invariant variables $\rh$, $p_{\rh}$, and not
to the gauge variant $\theta$, $p_{\theta}$. 
See the appendix for a description of the algorithm.
The random forces are
also not applied to $C$ and $P_C$ either, which makes it possible to monitor
thermalization of $P_C$. 
In practice the singularity of the polar coordinates can lead to problems
at the origin $\rho = 0$. We dealt with this in the following way.
The system is kicked at time intervals $\Delta t'=h$ by the random forces. 
If the kick is large in some sense, the updated coordinates may get too close 
to the origin, causing large discretization errors. 
We avoid this by replacing the single step update by a variable stepsize 
leapfrog 
algorithm for the integration of Hamilton's equations over the interval 
$(t',t'+h)$. The Langevin stepsize is still given by $h$, however.

The finite Langevin stepsize $h$ introduces an error which we
monitored by requiring the `output temperatures' $T'_{C}$ and
$T'_{\rh}$,
%obtained from quadratic terms in the hamiltonian,
\be
\lag \frac{\xi L P_C^2}{4\pi^2}\rag = T'_C = \beta^{-1}_C,\;\;\;\;
\lag \frac{p_{\rh}^2}{2a}\rag = T'_{\rh}=\beta^{-1}_{\rh}, \label{eq:Tout}
\ee
to be such that $\bt_{C,\rh}$ differed
less than 0.1 in absolute value from $\bt\equiv 1/T'$. This criterion is
based on the interpretation that small Langevin errors lead to an effective
temperature, the output temperature. Then the expectation
$\Gm'\propto\exp(-\bt E'_s)$, with $E'_s = 2\sqrt{2}/3 \approx 0.94$
suggests that an absolute error in $\bt$
of $\approx 0.1$, causes a relative error in $\Gm'$ of $\approx 10$\%.
%For low temperatures with $\bt\ageq 10$ this required $h\aleq ...$??

Having produced an independent configuration of $p$'s and $q$'s we
took this as the initial condition for the real time integration of
Hamilton's equations. For this we could use the original cartesian
coordinates, as the condition of charge zero is easily satisfied by
projecting regularly onto zero charge (this only involves changes of
machine precision order). After real time integration the Langevin
process was started again and the process was repeated untill
sufficient statistics was obtained.

Our configurations were actually microcanonical to some extent,
because the Langevin processes were stopped (for historical reasons)
at times that the total energy had its mean value. As expected, we
found in a few checks that the true canonical ensemble gave the same
results within errors. The friction parameter was taken $f=1$, and we
checked that results do not depend on $f$, as should be the case.

An accurate fifth order Adams-Bashforth-Moulton predictor-corrector
multi-step integration algorithm was used to keep the numerical drift
in the total energy sufficiently small. Otherwise we found that
the diffusion $\Dl$ turned out not to 
be linear in $t'$ for large $t'$.

\section{Low temperature region}

The results that will be presented are for $\xi = g^2/\lm =4$, for which
$m'_A = 2 = \sqrt{2}  m'_{\ph}$.
\begin{figure}[htb]
\epsfxsize 90mm
\centerline{\epsfbox{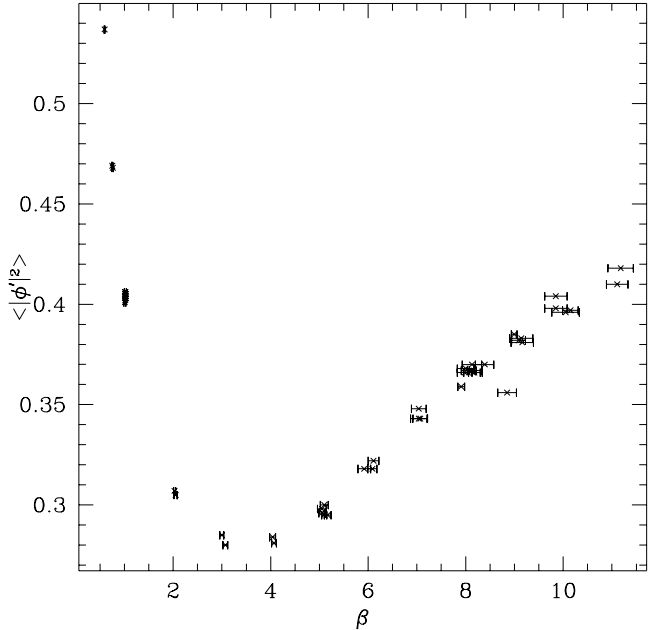}}
\caption[abh2]{ $\langle |\phi'|^2 \rangle$ versus $\beta$.
}
\label{fig:rho2}
\end{figure}
Fig.\ \ref{fig:rho2} shows the behavior of $\langle |\phi'|^2\rangle$ as a
function of $\bt$. We see a minimum near $\bt=3$ where the
crossover takes place between the low temperature region ($\bt > 3$)
and the high temperature region ($\bt < 3$). Recall that even
this `low temperature region' can be interpreted as a region of
high unscaled temperatures:
$T/m_{\ph} = v^2 /(\sqrt{2}\bt) \to \infty$ as $v^2 \to \infty$.

For low temperatures 
the Chern-Simons diffusion 
$\Dl$ tends to be small and large times $t$
were needed to get $\Dl > 1$.  For the evaluation of the rate
we require $\Dl > 10$,
such that at least three sphaleron transitions take place 
($\Dl C \ageq 3$).
%diffusion linear even for $C^2 > (N/2)^2$

\begin{figure}%[tbh]
\epsfxsize 140mm
\centerline{\epsfbox{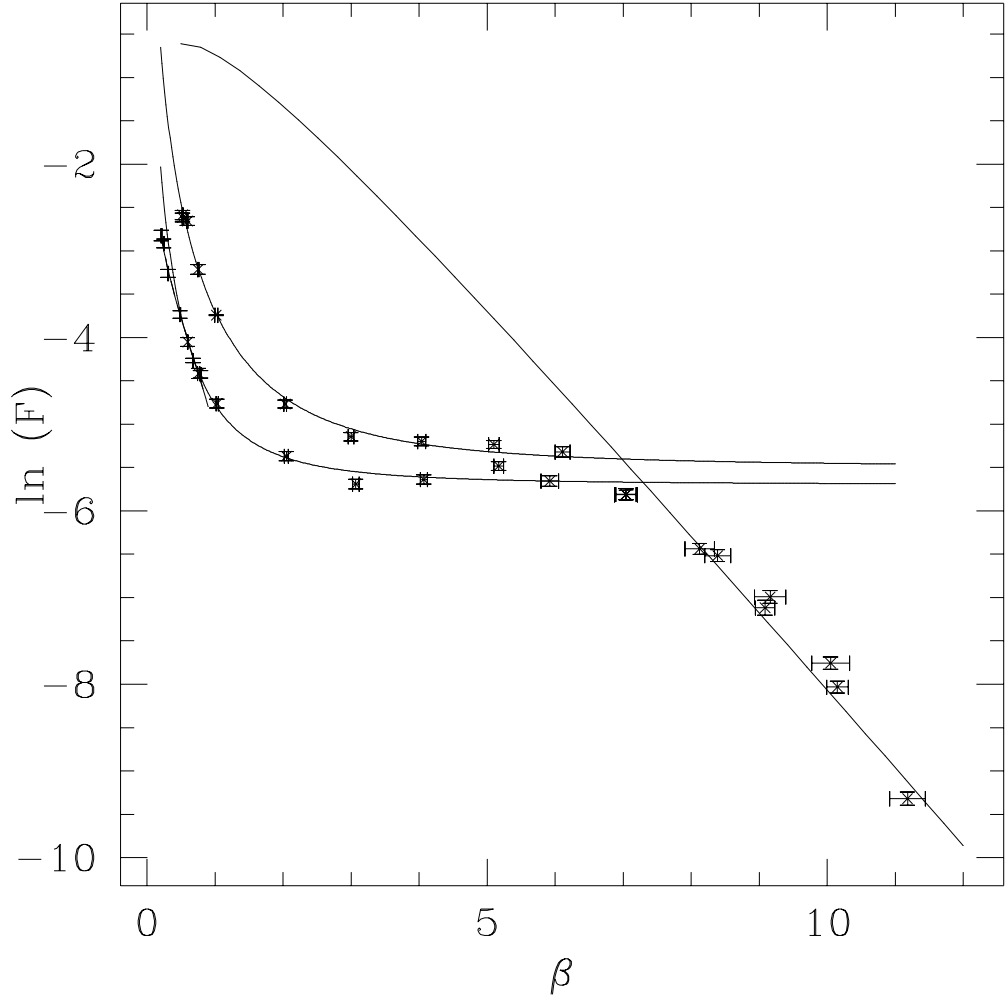}}
\caption[abh6]{Results for 
$\ln F\equiv\ln (\Gm/{m_{\ph}^2}L)$ for $\xi=4$ and volume
$L'=16$. The diagonal solid line represents the analytic semiclassical result
(\protect{\ref{eq:analy}}). The two solid lines extending from small
$\bt$ into the intermediate temperature region represent high 
temperature fits of the form $F = c_0 + c_2 \beta^{-2}$ 
\protect{\cite{SmTa95}}. 
The upper data are for $a'=0.32$, the lower data for $a'=0.16$.
The figure also shows at very small $\bt$ a fit of 
$F = c_0 + c_{2/3} \bt^{-2/3}$ to the $a'=16$ data.}
\label{fig:rate}
\end{figure}

In fig.~\ref{fig:rate} a
comparison is made with the analytic semiclassical result (\ref{eq:analy}).
The upper data are for
$a'=0.32$, the lower data for $a'=0.16$ (at $\beta=10,11$ for
$a'=0.32$ only). The system size is $L'=16$
($N=50$ and $N=100$). We obtained the same results within errors for
the smaller volume $L'=10.28$, in a check for a few $\bt$ values. 
However, for sizes as low as $L'=8$ we did see clear deviations in the low
temperature regime.
The errors in the rate $\Gm$ are statistical and determined with
the jackknife method from at least 900 configurations.
For $\bt$ we used the output temperature (\ref{eq:Tout}),
with errors obtained by the binning method.
The input $\bt$ is the nearest integer.

We conclude from fig.~\ref{fig:rate} that the classical simulation is
able to reproduce the semi-classical formula for $\beta \ageq 7$. For
the two lattice spacings $a' = 0.32$, 0.16 the classical rate in this
temperature region is lattice spacing independent, within the errors.
From the perturbative analysis in sect.\ 3 we see no reason to doubt
that the limit of zero lattice spacing exists in the weak coupling
region and that the resulting ratio
$\Gamma_{\mbox{class}}/\Gamma_{\mbox{semi-class}}$ approaches one as
$\beta \to \infty$.

\section{High temperature region}

\begin{figure}%[htb]
\centerline{\epsfxsize=70mm\epsfbox{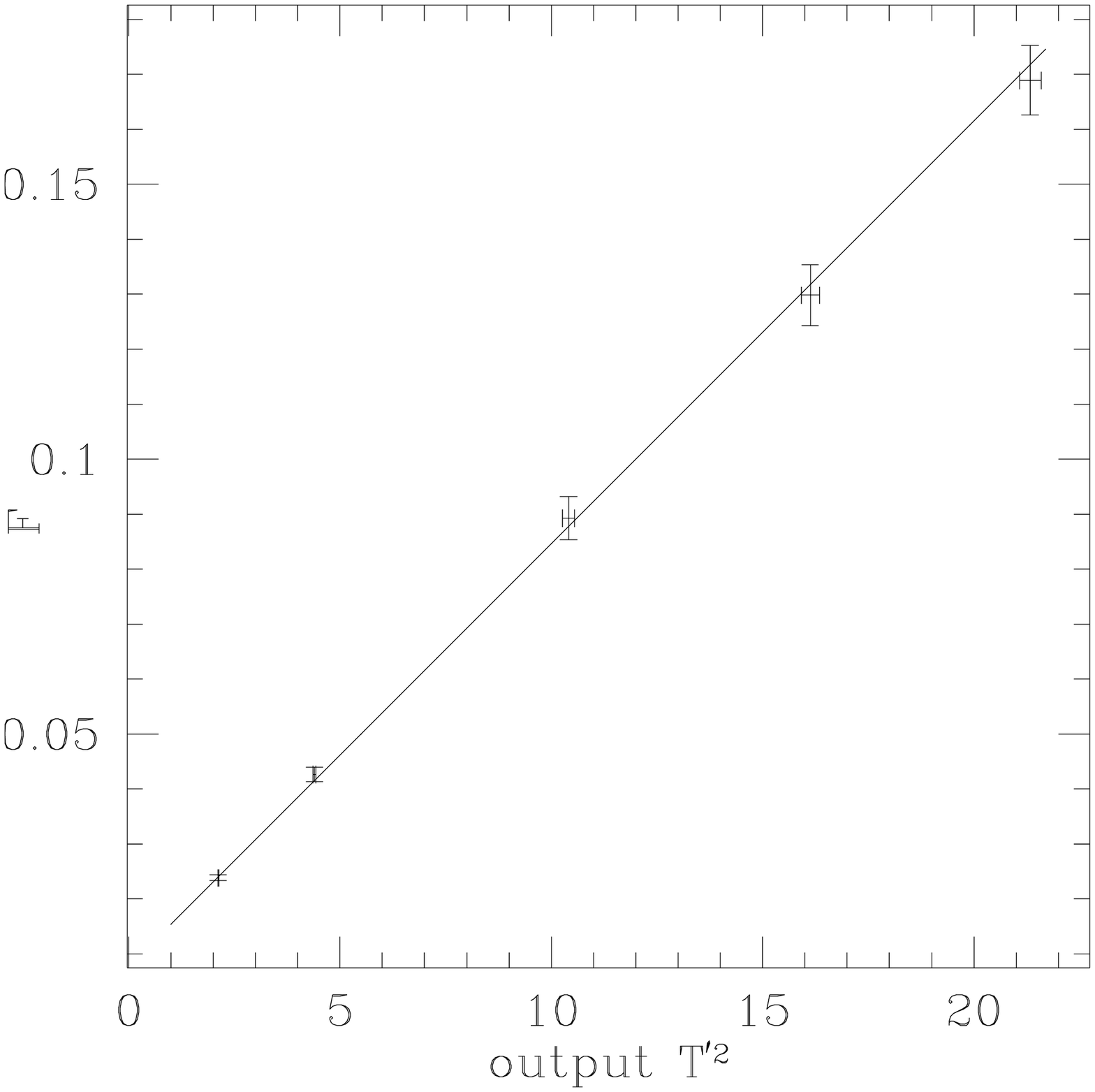}
            \epsfxsize=70mm\epsfbox{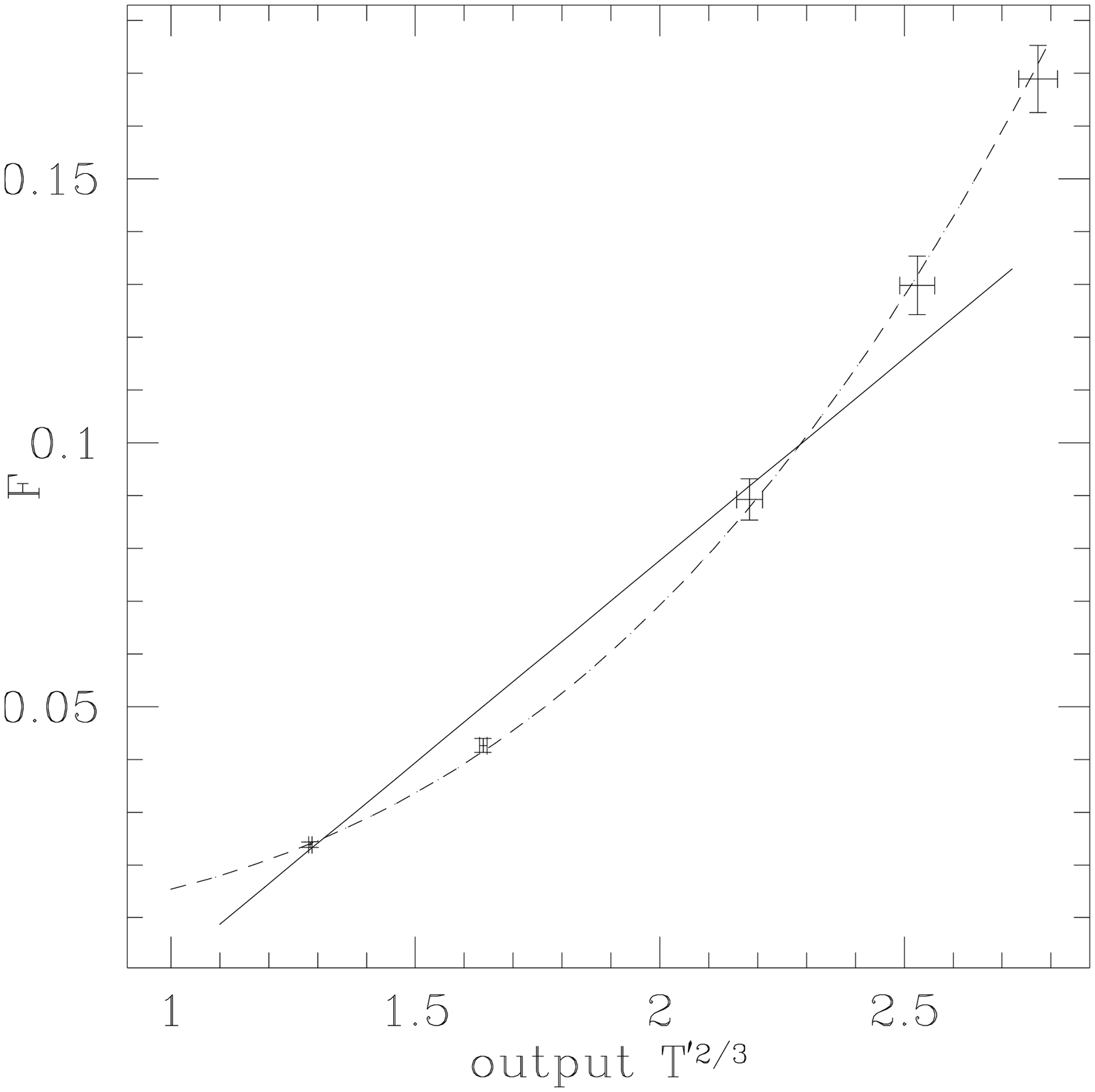}}
\caption[abh7]{Data showing $T^2$ behavior
for relatively large lattice spacing, 
$a' = 0.23$.
Left: $F$ versus $T'^2$ and a fit to $c_0 + c_2 T'^2$.
Right: same data versus $T'^{2/3}$ with the $T'^2$ fit of the left plot
(dashed line) and a fit to $c_0 + c_{2/3} T'^{2/3}$ (solid line). 
}
\label{fN71}
\end{figure}

\begin{figure}%[htb]
\centerline{\epsfxsize=70mm\epsfbox{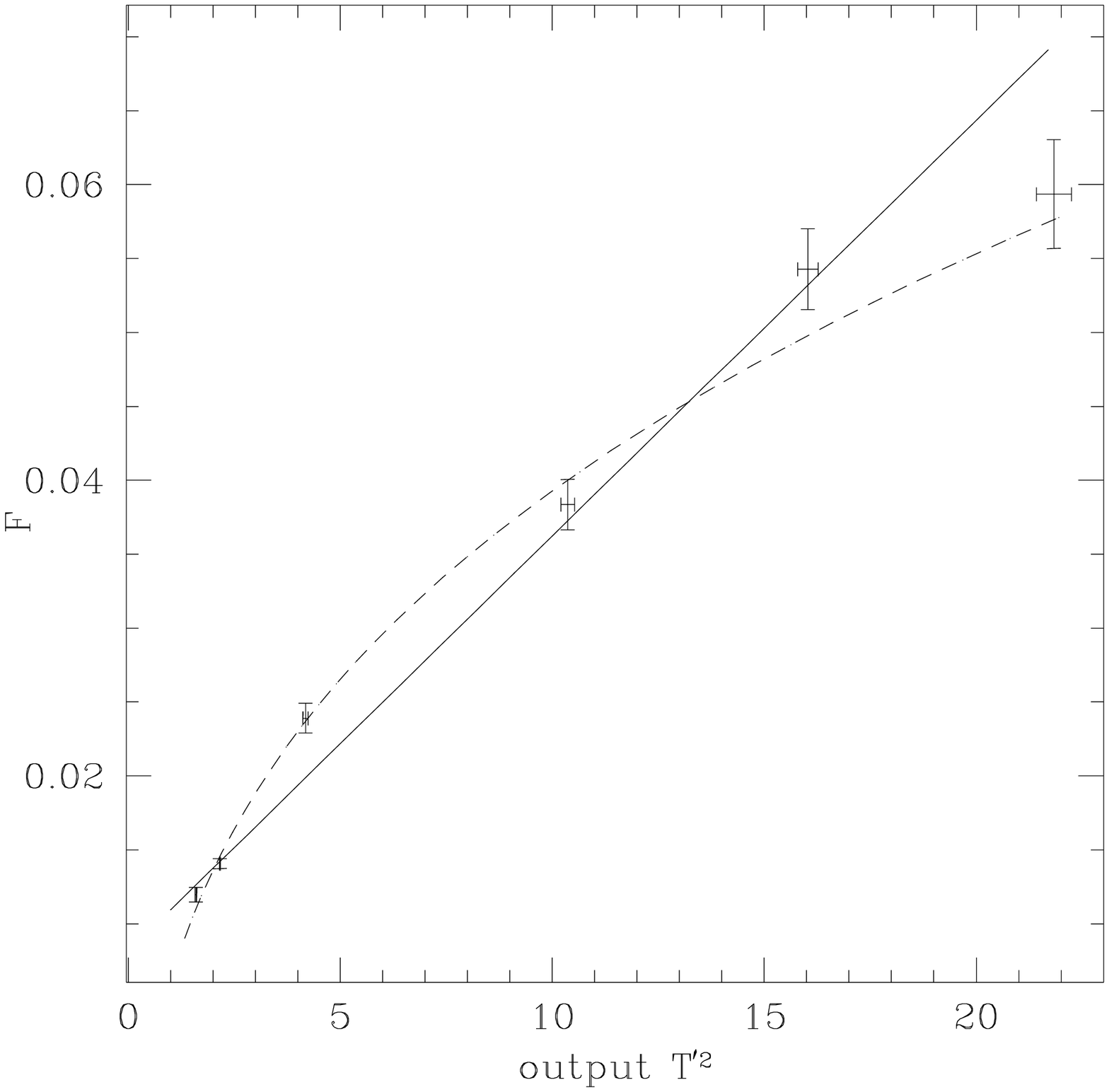}
            \epsfxsize=70mm\epsfbox{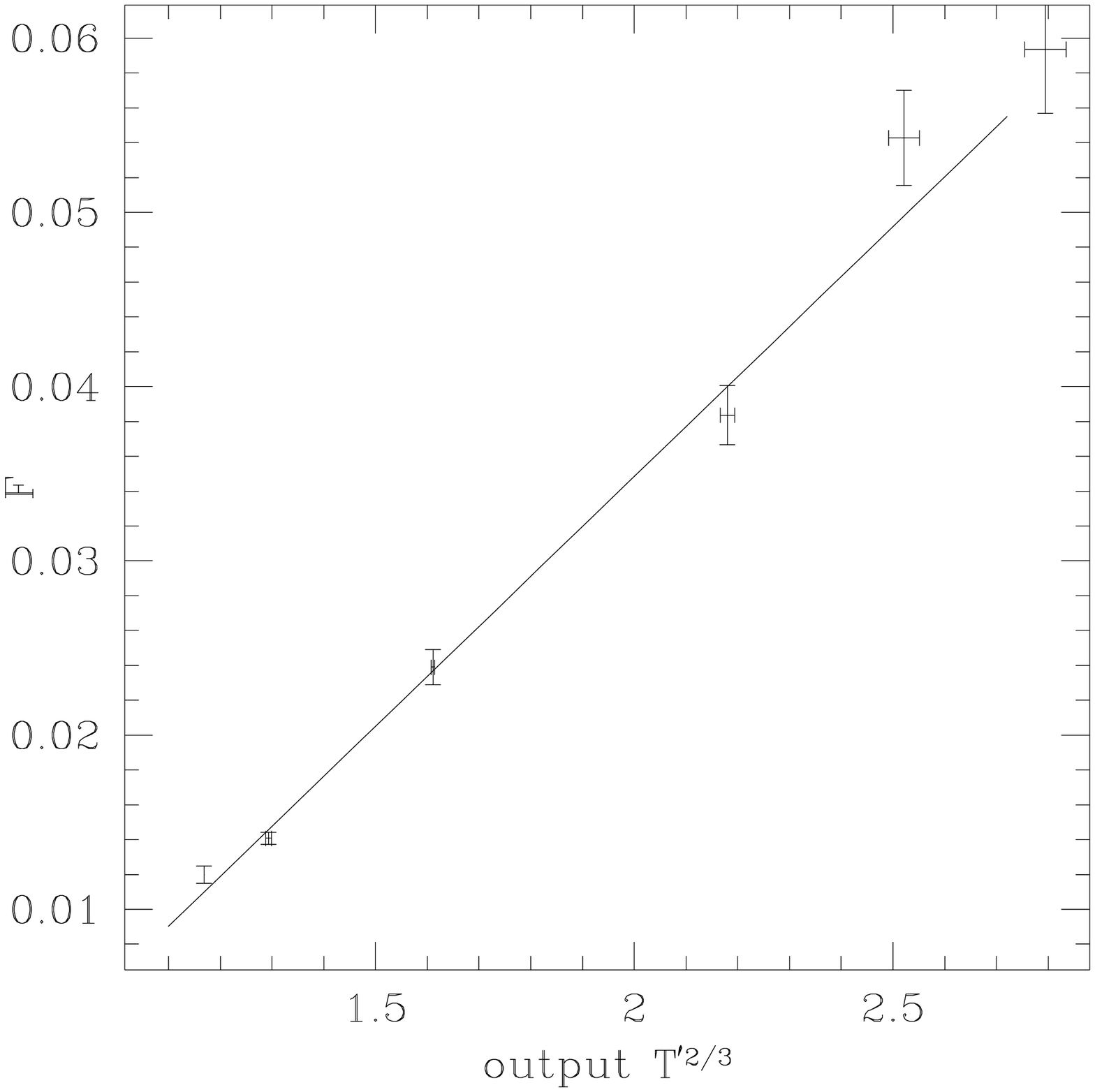}}
\caption[abh8]{Data illustrating the $T^{2/3}$ behavior
at small lattice spacings, $a'=0.16$.
Right: $F$ versus $T'^{2/3}$ and a fit to $c_0 + c_{2/3} T'^{2/3}$. 
Left: same data versus $T'^{2}$ with the $T'^{2/3}$ fit of the right plot
(dashed line), and a fit to $c_0 + c_{2} T'^{2}$ (solid line).
}
\label{fN100}
\end{figure}

\begin{figure}%[htb]
\epsfxsize 120mm
\centerline{\epsfbox{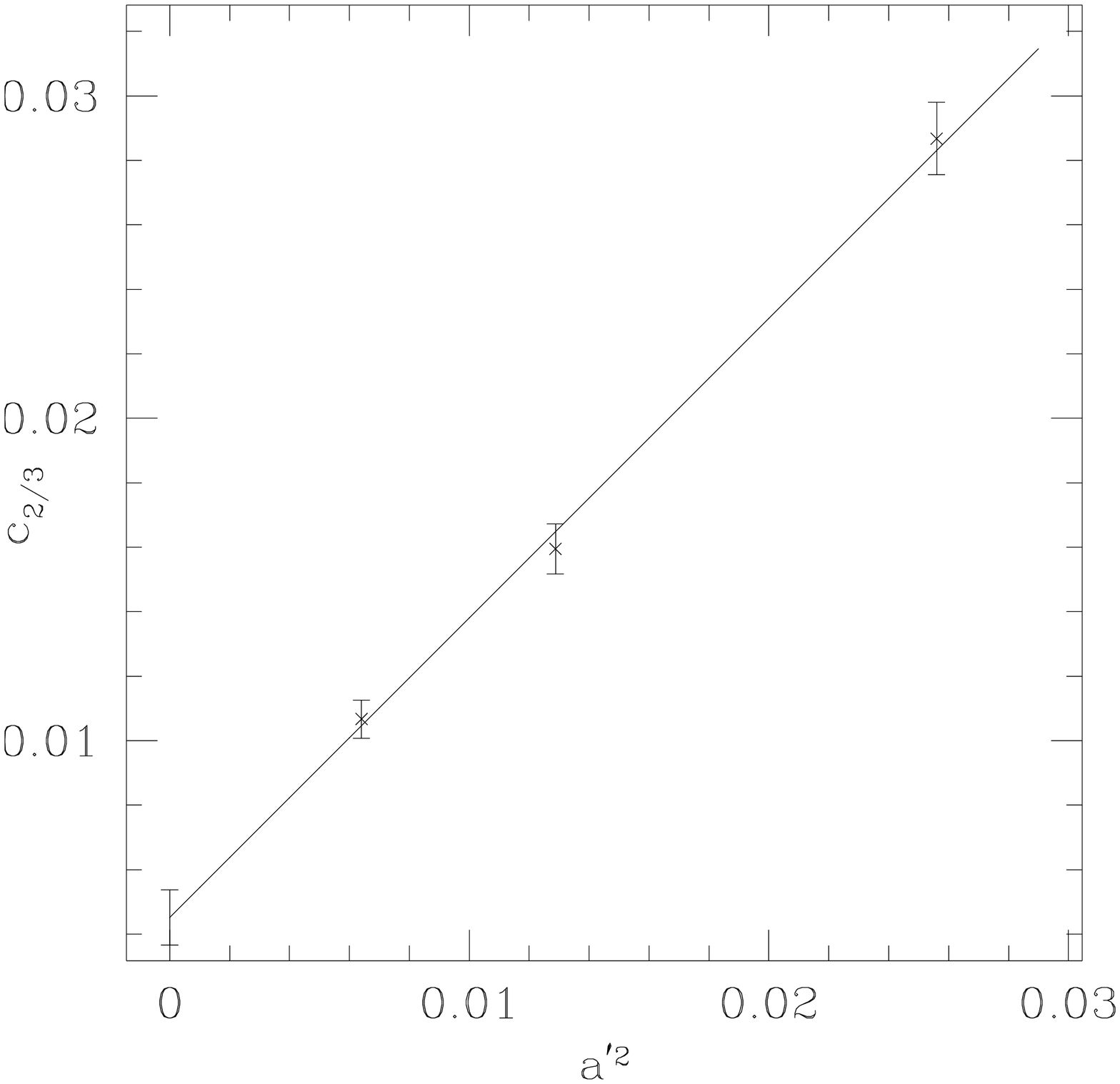}}
\caption[abh9]{Plot of $c_{2/3}$ as a function of $a'^2$.
}
\label{fato0}
\end{figure}

Decreasing $\beta$ we see in fig.\ \ref{fig:rate} that the rate is at first
more or less temperature-independent, after which an apparent $T^2$ behavior
sets in, which seems to confirm the expectation (\ref{Tsq}). 
However, there is substantial lattice spacing dependence in this temperature 
region and our earlier analysis lead to the conclusion that the coefficient
$c_2$ in the fit $F = c_0 + c_2 T'^2$ is proportional to $a'^2$,
such that the rate turned out
temperature-independent in the limit of zero lattice spacing \cite{SmTa95}. 
Here we reconsider the situation at higher temperatures.
We generated new data in the region $1.5 \leq T' \leq 5$ 
(input $\bt = 0.75$, 0.65, 0.45, 0.29, 0.23, 0.2), with volume $L' = 16$. 
At these temperatures
this value of $L'$ is amply sufficient for avoiding
significant volume dependence. 
The lattice spacings where given by $a' \approx 0.25$, 0.23, 0.16, 0.11, 0.08 
($a'= L'/N$, $N=63$, 71, 100, 141, 200). 
At first we tried extrapolating to zero lattice spacing and seeing if 
the resulting rate showed a $T^2$ or $T^{2/3}$ behavior, but in this way the 
errors turned out too large to allow for a 
meaningful discrimination between these two alternatives. 
We therefore looked at the
temperature dependence at fixed lattice spacing, with the following 
amusing outcome: the larger lattice spacings $a' = 0.25$, 0.23, 
favor $T^2$ behavior while the smaller spacings $a' = 0.16$, 0.11 and 0.08
favor $T^{2/3}$ behavior. Figs.\ \ref{fN71} and \ref{fN100} show two examples
(as before $T'$ in the plots is the output temperature).
Fitting the form $F = c_0 + c_2 T'^2$ (cf.\ (\ref{alternatives})) 
led to 
%
%$\chi^2/\mbox{d.o.f.} = 1.4$, 0.42, 8.6, 7.9, 3.45,
 $\chi^2/\mbox{d.o.f.} = 1.4$, 0.42, 6.5, 4.8, 3.45,
for $a' = 0.25$, 0.23, 0.16, 0.11, 0.08, respectively,
while $F = c_0 + c_{2/3} T'^{2/3}$ led to
%
%$\chi^2/\mbox{d.o.f.} = 28.9$, 21.6, 3.3, 1.8, 1.2.
 $\chi^2/\mbox{d.o.f.} = 28.9$, 21.6, 2.5, 1.1, 1.2.
Clearly,
the $T^{2/3}$ form is favored for the three smaller lattice spacings.
We extrapolated the resulting $c_{2/3}$ to zero lattice spacing, 
assuming a quadratic dependence on $a'$ (cf.\ fig.\ \ref{fato0}), with
the result $c_{2/3}(a'=0) = 0.00452 (86)$,
or 
\be 
\tilde\kp = 2 c_{2/3} = 0.0090(17),\;\;\;\;\; \xi=4.
\ee
It seems reasonable that the lattice spacing dependence %dies out 
falls
like $a^2$ as $a\to 0$, since the equations of
motion are based on an action with $O(a^2)$ discretization
errors. This could be upset by divergences, such that odd powers of
$a$ appear, as in 3+1 dimensions, but these are not expected here
since we have seen in sect.\ 3 that the classical theory is finite.

\section{Conclusion}
On the basis of perturbation theory we have argued that the effectively
classical approximation becomes exact in the limit (\ref{limit}) of weak 
coupling and high temperature. Furthermore, 
for correlation functions
the classical theory is finite,
which means that we can in principle obtain continuum answers by
extrapolating the numerical results to zero lattice spacing. This avoids
having to determine an effective action with lattice spacing dependent
counterterms, as we previously deemed necessary \cite{SmTa95}. 
In the limit (\ref{limit}) the purely
classical action is sufficient to leading order. 

In the resulting classical model we can still 
distinguish a low and high temperature regions separated by a crossover domain,
as illustrated by the behavior of the rate in Fig.\ \ref{fig:rate}.
In the low temperature region the rate is well described by the analytic
form based on the sphaleron \cite{BochTsi89}, as observed earlier
\cite{GriRuSha89,KraPo94,SmTa94}. The lattice spacing dependence is small
in this region, of the order of the error bars ($\approx 10\%$). 
Upon crossing over to the high temperature region
the lattice spacing dependence grows, and leads to a tricky phenomenon: 
apparently $\Gm/L \propto T^2$ at intermediate lattice spacings, but this 
turns into a $T^{2/3}$ behavior in the continuum limit. We obtained 
the estimate $\Gm/L = 0.0090(17) (\lm T)^{2/3}$, $T\to \infty$, for
$\xi = g^2/\lm = 4$. 

As mentioned in the Introduction, arguments for
a $T^{2/3}$ behavior were previously given in ref.\ \cite{FoKraPo94}. 
It was shown there that the classical rate could be written in
terms of a function $g(x,y)$, as 
$\Gm/L = T^{2/3} g(a^3 T, v^3/T)$,
using units in which $\lm = 1$, suppressing $\xi$ dependence. 
For high temperatures  $T^{2/3}$ behavior then followed 
in the continuum limit provided $g(0,0) \neq 0, \infty$ existed.
For the case $v^2=0$, numerical data supported a nontrivial $g(0,0)$.  
%One can argue that $v^2=0$ is able to exhibit
%the high temperature behavior of the $v^2 \neq 0$ case more easily. 
However, for $v^2=1$ (which corresponds to our scaled units (\ref{prime}))
the results for the rate were not compared to $T^{2/3}$ behavior with
much precision. The data looked compatible with our's at the time 
\cite{SmTa95}, but since it was taken at different $\xi$ ($\xi = 10$)
a detailed comparison was not meaningful. 
Our analytic results in Sect.\ 3 support the existence of a
continuum limit of $g(a^3 T,v^3/T)$, but notice how the 
combination $a^3 T$ implies 
non-commutativity of the limits $a\to 0$ and $T\to\infty$. Such
non-uniform behavior complicated our numerical analysis indeed.\\

\noindent

Acknowledgements\\
We like to thank A.I.~Bochkarev, A.~Krasnitz and G.~Aarts for useful 
discussions. This work is supported by FOM/NWO. Numerical simulations
were carried out on the CRAY C90 with financial support from NCF/NWO.

\appendix
\section{Simulation algorithm using Kramers equation}
This appendix we describe the second order
Langevin algorithm, closely related to `Kramers equation'.
Consider the class of systems described by the lagrangian
\be
L = \half g_{\al\bt}(q) \dot q^{\al} \dot q^{\bt} - V(q),
\ee
where $g_{\al\bt}$ plays the role of a metric in coordinate
space. We shall assume that $g_{\al\bt}$ can be expressed in the form
\be
g_{\al\bt} = e_{\al k} e_{\bt k},
\;\;\; \al, \bt, k = 1, \ldots M,
\ee
where $M$ is the number of $q$'s or $p$'s. 
The abelian Higgs model in polar coordinates can be described
this way. It is useful to have in mind a simple two dimensional
illustration given by 
\bea
L &=& \dot \phi^* \dot \phi - V(\phi)\\
&=& \dot\rho^2 + \rho^2 \dot\theta^2 - V(\rho,\theta),
\;\;\;\phi = \rho e^{i\theta},
\eea
for which 
\be
e_{1 1} = \sqrt{2},\;\;\;\;
e_{2 2} = \sqrt{2}\,\rho,
\;\;\;\;
e_{12}=e_{21}=0.
\ee
The hamiltonian is given by
\be
H = \half g^{\al\bt} p_{\al} p_{\bt} + V,
\;\;\;\;
p_{\al} = g^{\al\bt} \dot q_{\bt},
\ee
where $g^{\al\bt}$ is the inverse of $g_{\al\bt}$, $g^{\al\bt}
g_{\bt\gm} = \dl^{\al}_{\gm}$. 

Discretizing time with stepsize $h$, $t = nh$ ($n=0,1, \ldots$), 
we can derive a leapfrog algorithm by requiring the action
\be
S = \sum_n [p_{\al n} (q^{\al}_{n+1} - q^{\al}_n) - hH(p_n,q_n)]
\ee
to be stationary under variations of the canonical variables,
\bea
q_{n+1}^{\al} &=& q_n^{\al} 
+ h \frac{\partial H(p_n,q_n)}{\partial p_{\al n}},\\
p_{\al n+1} &=& p_{\al n} 
- h \frac{\partial H(p_{n+1},q_{n+1})}{\partial q^{\al}_{n+1}}.
\eea
We modify these equations into second order Langevin form
by adding friction and noise terms to the momentum equations,
\be
p_{\al n+1} &=& p_{\al n} 
- h \frac{\partial H(p_{n+1},q_{n+1})}{\partial q^{\al}_{n+1}}
-h F_{\al}^{\bt}(q_{n+1})p_{\bt n} \nonumber\\
& &+ \sqrt{2h}\, N_{\al k}(q_{n+1}) \nu_{k n},
\ee
where the $\nu_{k n}$ are normalized gaussian random numbers,
\be
\langle \nu_{k n} \rangle = 0,
\;\;\;\;
\langle \nu_{k n} \nu_{k' n'} \rangle = \dl_{k k'} \dl_{n n'}.
\ee
By a suitable choice of the random forces and friction the
system may equilibrate (at large times and in the limit of zero
stepsize) 
according to the Boltzmann distribution $\exp (-H/T)$.
For this it is necessary that
the Boltzmann distribution is a fixed point of the associated 
Fokker-Planck-type equation, which is called Kramers equation \cite{Ka}. 
This `stability condition'
leads to relations between $F$ and $N$, which will now be derived.

By the usual arguments one can derive the evolution equation
for the probability
distribution $W_t(p,q)$ at time $t$
(`Kramers equation') 
\bea
%\dot W &=& \lim_{h \to 0} \frac{W_{n+1} - W_n}{h}\\
\frac{W_{t+h} - W_t}{h}
&=&
\left(-
\frac{\partial H}{\partial p_{\al}}\, 
\frac{\partial}{\partial q^{\al}}
+
\frac{\partial H}{\partial q^{\al}}\, 
\frac{\partial}{\partial p_{\al}}
\right.\\ &&\mbox{}\left.
+
F_{\al}^{\al} + F_{\al}^{\bt} p_{\bt} \frac{\partial}{\partial p_{\al}}
+ M_{\al\bt} \frac{\partial^2}{\partial p_{\al}\partial p_{\bt}}
+ O(h)
\right )
W_t,
\eea
where
\be
M_{\al\bt} =  N_{\al k} N_{\bt k} .
\ee
The left hand side approaches $\dot W$ in the limit of zero stepsize and the
first two terms on the right hand side represent the usual 
Liouville flow, for which $\exp(-H/T)$ is a fixed point.
For the other terms $\exp(-H/T)$ is also a fixed point provided that
\be
0 = F_{\al}^{\al} - F_{\al}^{\bt} p_{\bt} 
\frac{\partial H}{T\partial p_{\al}}
+ M_{\al\bt}
\left(
\frac{\partial H}{T\partial p_{\al}}
\frac{\partial H}{T\partial p_{\bt}}
- \frac{\partial^2 H}{T\partial p_{\al}\partial p_{\bt}}
\right).
\ee
Inserting the explicit form of $H$ and comparing powers of the momenta
leads to the relations between $F$ and $N$,
\bea
F_{\al}^{\al} &=& \frac{1}{T}\, M_{\al\bt} g^{\al\bt},\\
\half(F_{\gm}^{\al} g^{\gm\bt} + F_{\gm}^{\bt} g^{\gm\al})
&=& \frac{1}{T} M_{\gm\dl} g^{\gm\al} g^{\dl\bt}.
\eea
This is satisfied by 
\be
F_{\al}^{\bt} = \frac{1}{T}\, M_{\al\gm} g^{\gm\bt},
\ee
which gives $F$ in terms of $M$ and the metric,
 
Ergodicity and stability should guarantee
that the evolution approaches the canonical distribution for $t\to\infty$.
This can be understood intuitively by writing
\be
W_t = e^{-H/2T}\, \tilde W_t.
\ee
The distribution $\tilde W_t$ evolves according to
\be
\frac{\tilde W_{t+h} - \tilde W_t}{h} +
\left(
\frac{\partial H}{\partial p_{\al}}\, 
\frac{\partial}{\partial q^{\al}}
-
\frac{\partial H}{\partial q^{\al}}\, 
\frac{\partial}{\partial p_{\al}}
\right) \tilde W_t
=
- P \tilde W_t,
\ee
with
\be 
P =
\left(-\frac{\partial}{\partial p_{\al}} + \frac{g^{\al\gm} p_{\gm}}{2T}\right)
M_{\al\bt}
\left(\frac{\partial}{\partial p_{\bt}} + \frac{g^{\bt\dl} p_{\dl}}{2T}\right).
\ee
The Liouville operator on the left hand side conserves probability.
On the other hand, since $M$ is evidently a positive matrix, 
the differential operator $P$ is positive definite. Expanding $\tilde W_t$ in
terms of the eigenmodes of $P$, the modes with nonzero eigenvalues are expected
to die out during the time evolution, such that $\tilde W_t$ approaches
the zero mode $\exp (-H/2T)$,
\be
\left(\frac{\partial}{\partial p_{\bt}} + \frac{g^{\bt\dl} p_{\dl}}{2T}\right)
e^{-H/2T} = 0,
\ee
which is of course also a zero mode of the Liouville operator.
The original $W_t$ then approaches the Boltzmann distribution with
temperature $T$.

A natural choice for $N$ is given by
\be
N_{\al k} = \sqrt{fT}\, e_{\al k},
\;\;\;\;
M_{\al \bt} = fT g_{\al \bt},
\;\;\;\;
F_{\al}^{\bt} = f \dl_{\al}^{\bt},
\ee
with some arbitrary friction coefficient $f>0$.
However, in our application to the abelian Higgs model 
we want to add noise only to the radial momenta $p_{\rh}$
and not to the angular momenta $p_{\theta}$. Therefore,
$N$ and $M$ contain a projector onto the radial variables,
\be
N_{\al k} = \sqrt{fT}\,e_{\al l} \pi_{lk},
\;\;\;\;
\pi_{kl}\pi_{lm} = \pi_{km},
\ee
where $\pi$ is the projector.
In the above two dimensional model this is illustrated by
\be
\pi_{11} = 1,\;\;\;\; \pi_{12} = \pi_{21} = \pi_{22} = 0,
\ee
and consequently
\be
N_{1 1} =  \sqrt{2fT},
\;\;\;\;
M_{1 1} = 2fT,
\;\;\;\;
F_{1}^{1} = f,
\ee
with the other components vanishing. These $N$ and $F$ are independent
of $\rho$ and $\theta$.


\begin{thebibliography}{99}
\bibitem{CoKaNe93} A.G. Cohen, D.B. Kaplan, A.E.~Nelson,
                   Ann.\ Rev.\ Nucl.\ Part.\ Sci.\ 43 (1993) 27.
\bibitem{RuSha96}  V.A.~Rubakov and M.E.~Shaposhnikov, 
                   Usp.\ Fiz.\ Nauk 166i (1996) 493, Phys.\ Usp.\ 39 (1996) 461,
                   hep-ph/9603208.
\bibitem{GriRu88}   D.Yu.~Grigoriev and V.A.~Rubakov, 
                   Nucl.\ Phys.\ B299 (1988) 67.
\bibitem{GriRuSha89}  D.Yu.~Grigoriev, V.A.~Rubakov and M.E.~Shaposhnikov,
                   Nucl.\ Phys.\ B326 (1989) 737.
\bibitem{BochFo91}   A.I.~Bochkarev and Ph.~de Forcrand, 
                   Phys.\ Rev.\ D44 (1991) 519.
\bibitem{SmTa94}   J.~Smit and W.H.~Tang, 
                   Nucl.\ Phys.\ B (Proc.\ Suppl.) 34 (1994) 616.
\bibitem{KraPo94}  A.~Krasnitz and R.~Potting, 
                   Nucl.\ Phys.\ B (Proc.\ Suppl.) 34 (1994) 613.
\bibitem{FoKraPo94}P.~de Forcrand, A.~Krasnitz and R.~Potting, 
                   Phys.\ Rev.\ D50 (1994) 6054.
\bibitem{SmTa95}   J.~Smit and W.H.~Tang, 
                   Nucl.\ Phys.\ B (Proc.\ Suppl.) 42 (1995) 590.
\bibitem{BochTsi89} A.I.\ Bochkarev and G.G.\ Tsitsishvili, 
                   Phys.\ Rev.\  D40 (1989) 1378.
\bibitem{Ambea}    J.~Ambj\o rn, T.~Askgaard, H.~Porter and M.E.~Shaposhnikov,
                   Phys.\ Lett.\ B244 (1990) 479; Nucl.\ Phys.\ B353 (1991) 346;
\bibitem{AmKra95}  J.~Ambj\o rn and A.~Krasnitz, Phys.\ Lett.\ B362 (1995) 97;
                   Nucl.\ Phys.\ B506 (1997) 387.
\bibitem{TaSm96}    W.H.~Tang and J.~Smit, Nucl. Phys. B482 (1996) 265.
\bibitem{MoTu97}   G.D.~Moore and N.G.~Turok, Phys.\ Rev.\ D56 (1997) 6533.
\bibitem{MoHuMu97} G.D.~Moore, C.H.~Hu and B.~M\"uller, hep-ph/9710436.
\bibitem{Mo98}     G.D.~Moore, hep-ph/9801204.
\bibitem{BoMcLeSmil95} D.~B\"odeker, L.~McLerran and A.~Smilga,
                    Phys.\ Rev.\ D52 (1995) 4675.  
\bibitem{Ar97}      P.~Arnold, Phys.\ Rev.\ D 55 (1997) 7781.
\bibitem{AaSm96}    G.~Aarts and J.~Smit, Phys.\ Lett.\ B393 (1997) 395.
\bibitem{AaSm97}    G.~Aarts and J.~Smit, Nucl.\ Phys.\ B511 (1998) 451.
\bibitem{BuJa97}   W.~Buchm\"uller and A.~Jakovac,
                    Phys.\ Lett.\ B407 (1997) 39.
\bibitem{TaSm97}    W.H.~Tang and J.~Smit, Nucl.\ Phys.\ B510 (1998) 401.
\bibitem{BoLa97}    B.~B\"odeker and M.~Laine, hep-ph/9707489.
\bibitem{ArYa97}    P.~Arnold and L.G.~Yaffe, Phys.\ Rev.\ D57 (1998) 1178.
%\bibitem{Sm97}     J.~Smit, hep-lat/9708??.
\bibitem{JoKoSi79}   D.~Jones, J.~Kogut and D.~Sinclair, 
                   Phys.\ Rev.\ D10 (1979) 1882.
\bibitem{IchMa94}  I.~Ichinose, H.~Mukaida, 
                   Int.\ J.\ Mod.\ Phys.\ A9 (1994) 1043.
\bibitem{DiHe96}   H.~Dilger and J.~Heitger, 
                   Nucl.\ Phys.\ B (Proc.\ Suppl.) 53 (1997) 587.
\bibitem{KhlSha88} S.Yu.~Khlebnikov and M.E.~Shaposhnikov,
                   Nucl.\ Phys.\ B308 (1988) 885.
\bibitem{BochShap87}A.I.~Bochkarev and M.E.~Shaposhnikov,
                   Mod.\ Phys.\ Lett.\ A12 (1987) 991.
\bibitem{LeBe}     M.~Le Bellac, Thermal Field Theory, CUP 1996.
\bibitem{Ka}       N.G.~van Kampen, 
                   Stochastic Processes in Physics and Chemistry,
                   North-Holland 1981. 
\end{thebibliography}
\end{document}